\documentclass[journal=ancac3,manuscript=article,layout=twocolumn]{achemso}
\usepackage[utf8]{inputenc}
\usepackage{graphicx, subfigure, amsfonts,amssymb,amsmath, array, gensymb, chemformula, lipsum, caption, xcolor, gensymb}
\usepackage{hyperref}
\usepackage[T1]{fontenc}

\title{Domain-dependent surface adhesion in twisted few-layer graphene: Platform for moir\'e-assisted chemistry }

\author{Valerie Hsieh}
\altaffiliation{These authors contributed equally to this work}
\author{Dorri Halbertal}
\altaffiliation{These authors contributed equally to this work}
\author{Nathan R. Finney}
\affiliation[Columbia University]
{Department of Physics, Columbia University, New York, NY 10027, USA}
\author{Ziyan Zhu}
\affiliation[Harvard University]
{Department of Physics, Harvard University, Cambridge, MA 02138, USA}
\alsoaffiliation [SLAC] {Stanford Institute for Materials and Energy Sciences,
SLAC National Accelerator Laboratory, Menlo Park, CA 94025, USA}
\author{Eli Gerber}
\affiliation[Cornell University]
{School of Applied and Engineering Physics, Cornell University, Ithaca, NY 14853, USA}
\author{Michele Pizzochero}
\affiliation[SEAS Harvard University]
{John A. Paulson School of Engineering and Applied Sciences,
Harvard University, Cambridge, MA 02138, USA}

\author{Emine Kucukbenli}
\affiliation[SEAS Harvard University]
{John A. Paulson School of Engineering and Applied Sciences,
Harvard University, Cambridge, MA 02138, USA}
\alsoaffiliation[Boston University]{Information Systems Department,
Boston University, Boston, MA 02215, USA}
\author{Gabriel R. Schleder}
\affiliation[SEAS Harvard University]
{John A. Paulson School of Engineering and Applied Sciences,
Harvard University, Cambridge, MA 02138, USA}
\alsoaffiliation[LNNano]{Brazilian Nanotechnology National Laboratory, CNPEM, 13083-970 Campinas, São Paulo, Brazil}
\author{Mattia Angeli}
\affiliation[SEAS Harvard University]
{John A. Paulson School of Engineering and Applied Sciences,
Harvard University, Cambridge, MA 02138, USA}

\author{Kenji Watanabe}
\affiliation[NIMS]{Research Center for Functional Materials, National Institute for Materials Science, 1-1
Namiki, Tsukuba 305-0044, Japan}
\author{Takashi Taniguchi}
\affiliation[NIMS]
{International Center for Materials Nanoarchitectonics, National Institute for Materials
Science, 1-1 Namiki, Tsukuba 305-0044, Japan}
\author{Eun-Ah Kim}
\affiliation[Cornell University]
{Department of Physics, Cornell University, Ithaca, NY 14853, USA}
\author{Efthimios Kaxiras}
\affiliation[Harvard University]
{Department of Physics, Harvard University, Cambridge, MA 02138, USA}
\alsoaffiliation[SEAS Harvard University]
{John A. Paulson School of Engineering and Applied Sciences,
Harvard University, Cambridge, MA 02138, USA}
\author{James Hone}
\author{Cory R. Dean}
\email{cd2478@columbia.edu}
\author{D. N. Basov}
\affiliation[Columbia University]
{Department of Physics, Columbia University, New York, NY 10027, USA}
\email{db3056@columbia.edu}

\begin{document}

\maketitle 
\twocolumn[
\begin{@twocolumnfalse}
\section{Abstract}
Twisted van der Waals multilayers are widely regarded as a rich platform to access novel electronic phases, thanks to the multiple degrees of freedom available for controlling their electronic and chemical properties. Here, we propose that the stacking domains that form naturally due to the relative twist between successive layers act as an additional "knob" for controlling the behavior of these systems, and report the emergence and engineering of stacking domain-dependent surface chemistry in twisted few-layer graphene. Using mid-infrared near-field optical microscopy and atomic force microscopy, we observe a selective adhesion of metallic nanoparticles and liquid water at the domains with rhombohedral stacking configurations of minimally twisted double bi- and tri-layer graphene. Furthermore, we demonstrate that the manipulation of nanoparticles located at certain stacking domains can locally reconfigure the moir\'e superlattice in their vicinity at the $\micro$m-scale. Our findings establish a new approach to controlling moir\'e-assisted chemistry and nanoengineering.

\smallskip
\smallskip
keywords: twisted graphene moir\'e, rhombohedral and Bernal stacking domains, nanoengineering, surface chemistry

\end{@twocolumnfalse}
]

\clearpage
\newpage
%%% FIG 1
\begin{figure*}[h!]
    \centering
    \includegraphics[width=14.86cm]{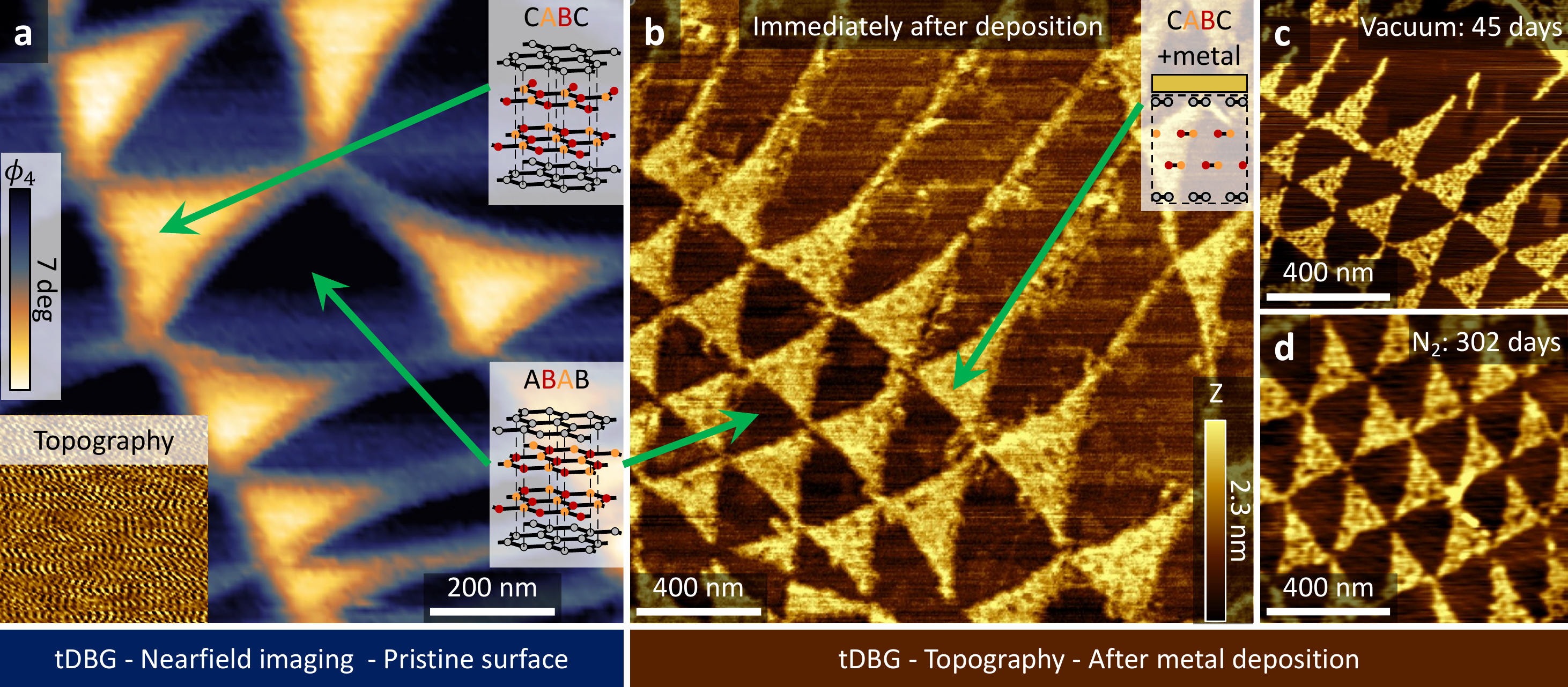}
    \caption{{\textbf{Moir\'e-assisted chemistry in twisted double bilayer graphene} \textbf{(a)} Mid-IR nearfield imaging of the twisted double bilayer moir\'e superlattice (tDBG) moir\'e superlattice (phase, 4th harmonic). Image taken on pristine sample before deposition. No topographic features were observed (see simultaneously taken tapping mode scan in inset). The Bernal (ABAB) and rhombohedral (ABCA) domain configurations are illustrated over the image. \textbf{(b)} Metal nano-structures forming over the rhombohedral phase of tDBG, revealed by tapping mode atomic force microscopy (AFM). The patterns formed after mechanical ablation deposition of Field's metal over a minimally twisted open phase tDBG (see Supporting Note A).  \textbf{(c)} The same measurement performed after 45 days during which the sample was stored in vacuum showing degradation of fine features. \textbf{(d)} Storing the sample in dry N$_2$ at room temperature stopped further degradation, even after more than 300 days. All tapping mode scans share the same color-bar.} \label{Fig: selective_adhesion_TDBG}}
\end{figure*}

%\section{Introduction}

Twisted van der Waals (vdW) heterostructures have drawn growing interest in recent years~\cite{geim2013van,Bistritzer2011,MacDonald2020,Kennes2021,Carr2020,Lau2022,Tritsaris2021}. 
The twist angle creates a moir\'e pattern at a larger length scale than the periodicity in each independent layer, leading to spatially-dependent interactions between the layers.
As such, the twist angle has been used as a continuous knob to effectively tune the Hamiltonian of the system and realize novel electronic phases~\cite{Cao2018, Cao2018_2}. At small twist angles, large domains of lowest energy stacking configurations separated by sharp domain walls form through the atomic relaxation process\cite{carr:2018:relaxation, Halbertal2021, Yoo2019,Wong2015,Huang2018}, resulting in multiple different stacking orders coexisting within the multilayer.

Recently, several works have explored how the twist angle and the consequent changes to the global electronic structure of the material, such as the emergence of flat bands or modified density of states, affect the surface chemistry and catalytic properties of moir\'e systems \cite{Kwabena2022NatureChemistry,Ding2016,Ding2018,LiangFu2021_moiremetal}. 
Additionally, preferential adsorption of chemical species due to corrugation and strain effects on the moir\'e surface has also been observed \cite{Lu2012_C60}. In general, strain has been utilized as a means of manipulating the moir\'e patterns in twisted vdW heterostructures~\cite{Gao2021heterostrain,qiao2018,huder2018}.
However, until now, no investigation has focused on the local stacking order as a platform for controlling the surface chemistry of this class of materials.

Here, we demonstrate domain-dependent surface adhesion as a new avenue to observe and manipulate the surface chemistry of moir\'e systems, a phenomenon we call moir\'e-assisted chemistry. We first show selective adhesion of metallic film on the rhombohedral domain of twisted double bilayer graphene (tDBG) while avoiding the Bernal domain. We then explore the robustness of this observation in twisted double trilayer graphene (tDTG) under varying ambient conditions and through domain-dependent adhesion of another material, namely, water droplets. Finally, we demonstrate how the particles that selectively adhere to the surface can be used to reshape the underlying moir\'e lattice. Overall, our results establish a basis for the development of capabilities in moir\'e engineering via domain-dependent surface interactions.

% \section{Results and Discussion}

In tDBG, the relaxed moir\'e superlattice is composed of two triangular stacking order domains: the lowest energy Bernal configuration (ABAB) and the metastable rhombohedral configuration (ABCA) \cite{Halbertal2021,Kerelsky2021}. Due to the small but finite energy difference between the Bernal phase and the rhombohedral phase, the rhombohedral domain walls exhibit a finite inward curvature~\cite{Halbertal2021}. This characteristic curving allows us to unambiguously identify the rhombohedral domains with scanning probe microscopy techniques (see Supporting Note A). Fig. \ref{Fig: selective_adhesion_TDBG}a shows a mid-IR scanning nearfield optical microscopy (mIR-SNOM) map measured on a tDBG sample with a 0.3-degree twist angle. The rhombohedral and Bernal stacking domains are identified by the light and dark contrast respectively (see Fig. \ref{Fig: selective_adhesion_TDBG}a insets); simultaneous topography imaging (lower left inset of Fig. \ref{Fig: selective_adhesion_TDBG}a) shows that the sample is featureless on the surface. We deposit Field's metal onto this surface using a mechanical ablation technique (see Supporting Note A). This process breaks the molten metal and scatters its nanoparticles on the surface. One would naively expect the formation of random patches of metallic film on the surface, with no correlation to the moir\'e superlattice. Instead, the tapping mode atomic force microscopy (AFM) map of Fig. \ref{Fig: selective_adhesion_TDBG}b reveals the formation of metallic films with complete correlation to the rhombohedral domains of the moir\'e superlattice, as imaged in mid-IR nearfield. The triangular metallic film domains exhibit sharp corners and fine features, as narrow as the thin double domain walls separating two Bernal phases\cite{Halbertal2021}, as shown in Fig. \ref{Fig: selective_adhesion_TDBG}b (diagonal lines along the top right part of the panel). 

After 45 days of storage within vacuum, the same AFM measurement over the surface of the sample shows degradation of fine nanoparticle features. Particularly along double domain walls, previously visible nanoparticle features no longer appear in subsequent topography scans (Fig. \ref{Fig: selective_adhesion_TDBG}c). After 173 additional days and again 302 days of storage in a dry N$_2$ environment, no further degradation is observed (Fig. \ref{Fig: selective_adhesion_TDBG}d). Robust coverage of the larger triangular regions persists regardless of the storage conditions of the sample. 
%The degradation of fine features over time in vacuum and halting of further degradation via dry N$_2$ environment suggests the presence of an external, volatile agent that is critical for the observed phenomenon. Indeed, 
Theoretical investigations using first principles density-functional theory (DFT) calculations indicate that domain-dependent adhesion is not intrinsic to the tDBG system. The calculated energy difference between the Bernal and rhombohedral double bilayer graphene changes only slightly with the addition of the metal, on the order of a few meV/unit cell, depending on the details of the metal nanoparticle model (see Supporting Note B). This energy difference is too small to account for the robust experimental observations. We therefore suspect that there is likely a mediating agent that plays a critical role in the emergence of selectivity. We hypothesize that the mediating agent is most likely a carbon-based organic substance, used either in the mechanical exfoliation of our materials or in the fabrication of the sample (see Supporting Notes A and D).

To probe whether the previously seen selectivity is unique to the tDBG system, we investigate a different multilayered graphene system, tDTG. Like tDBG, the tDTG system has two nearly degenerate low-energy stacking configurations. 
We observe domain-dependent adhesion in tDTG as well, further supporting that the phenomenon is universal, with an apparent selectivity determined by stacking domain. As shown recently \cite{Halbertal_2022}, in tDTG, energy considerations lead to a global layer translation, resulting in the moir\'e superlattice being ABABAB/BCBACA (see Fig. S3a). The ABABAB is a pure Bernal stacking configuration and the BCBACA phase can be thought of as a 4-layered rhombohedral graphene encapsulated by two Bernal interfaces. 

%%% FIG 2
\begin{figure*}[h!]
    \centering
    \includegraphics[width=17cm]{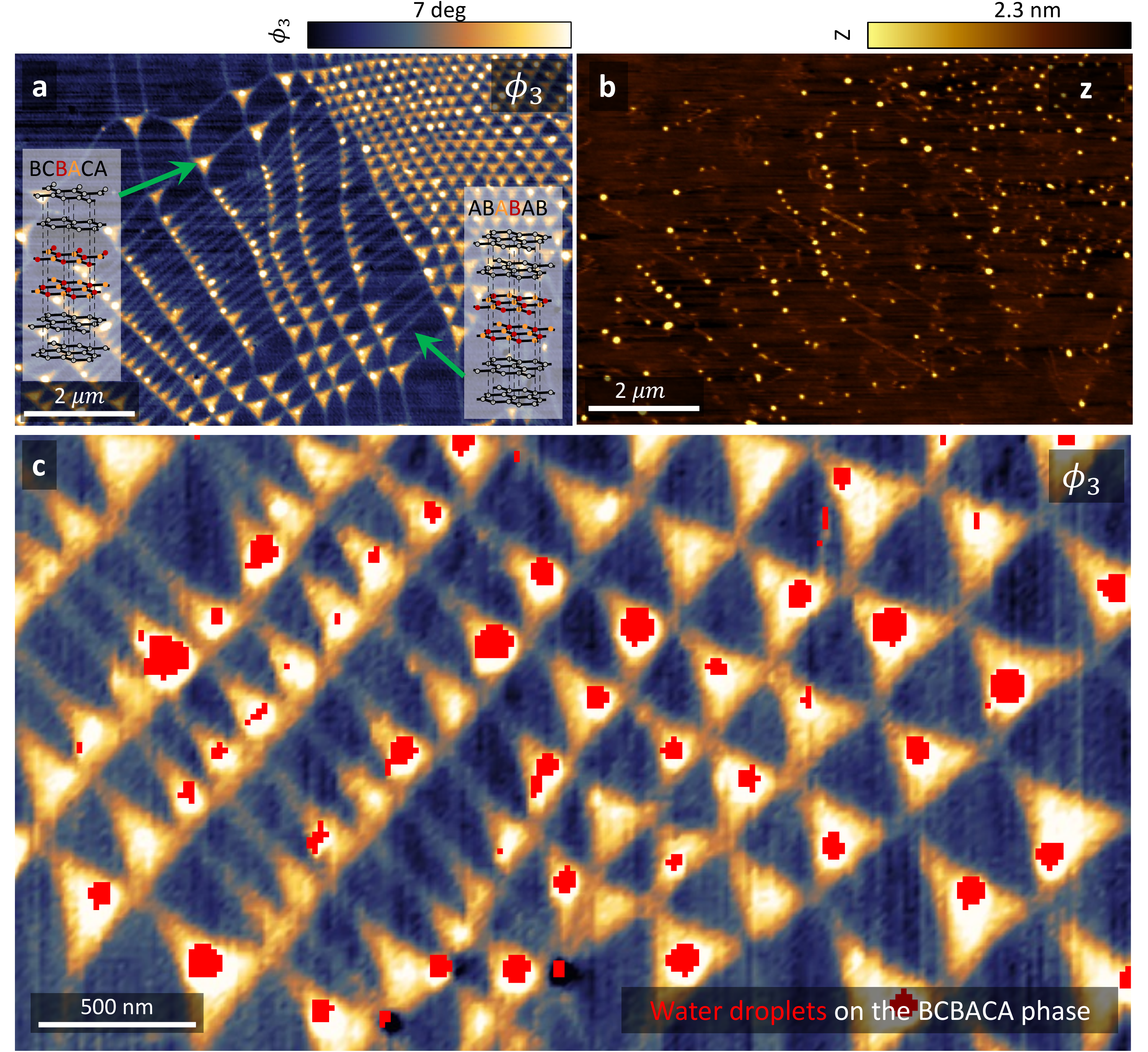}
    \caption{{\textbf{Moir\'e-assisted chemistry in twisted double trilayer graphene} \textbf{(a)} For small twist angles the twisted double trilayer graphene system (tDTG) forms domains of ABABAB and BCBACA stacking configurations \cite{Halbertal_2022}.  Mid-IR nearfield phase (3rd demodulation harmonic) of the tDTG, revealing the moir\'e superlattice. \textbf{(b)} A topography map of the moir\'e superlattice after wetting of the surface (see text for details), showing emergent topographic features. The scan was taken simultaneously with \textbf{(a)}. \textbf{(c)} Finer resolution nearfield phase map of the tDTG sample, with location of water droplets (from simultaneously acquired topographic map) overlayed in red. The correlation between water droplets and one phase of the moir\'e superlattice demonstrates moir\'e selective wetting.} \label{Fig: selective_adhesion_TDTG}}
\end{figure*}

Similar to tDBG, we investigate the tDTG sample using mid-IR nearfield imaging and verify the existence of moir\'e patterns with domains of BCBACA and Bernal stacking order (see Fig. S3b). A simultaneously taken tapping mode AFM scan (see Fig. S3c) reveals no topographic features corresponding to the moir\'e pattern. We then place the tDTG sample into a humid environment (see Supporting Note D). Upon removal, we repeat simultaneous mid-IR nearfield and tapping mode AFM scans. Remarkably, while mid-IR scans reveal the moir\'e are unchanged from exposure to humid conditions (Fig. \ref{Fig: selective_adhesion_TDTG}a), tapping AFM scans (scan of Fig. \ref{Fig: selective_adhesion_TDTG}b) show topographic features that correlate with the BCBACA phase of tDTG, as reflected in Fig. \ref{Fig: selective_adhesion_TDTG}c. The preferential adhesion of water molecules to the rhombohedral domain appears to be analogous to the tDBG system, albeit with qualitative differences. For instance, while the metallic film in tDBG completely covered the rhombohedral phase, the water molecules in tDTG form droplets which encompass the BCBACA phase, but do not extend to the domain boundary. 

We note that the domain-preferential adhesion of water in the tDTG sample is a transitory effect, as repeating the same experiment several months later does not reproduce the results. Our {\em ab initio} molecular dynamics and continuum solvation model calculations confirm that liquid water alone does not exhibit domain-dependent adhesion, and thus, it cannot serve as a sole mediating agent for domain dependence (see Supporting Note E). Although our theoretical calculations model focus on the double bilayer graphene system and not on the double trilayer graphene system, these calculations confirm that liquid water is too distant from the surface to experience an electrostatic difference between the rhombohedral and Bernal phases, a result which should be analogous in both tDBG and tDTG systems. The transitory nature of this phenomenon further indicates that it is not intrinsic to the studied system. A comprehensive description of the history of a tDBG sample displaying off-to-on selectivity of Field's metal adhesion is provided in Supporting Note F. 

\begin{figure*}[h!]
    \centering
    \includegraphics[width=15.24cm]{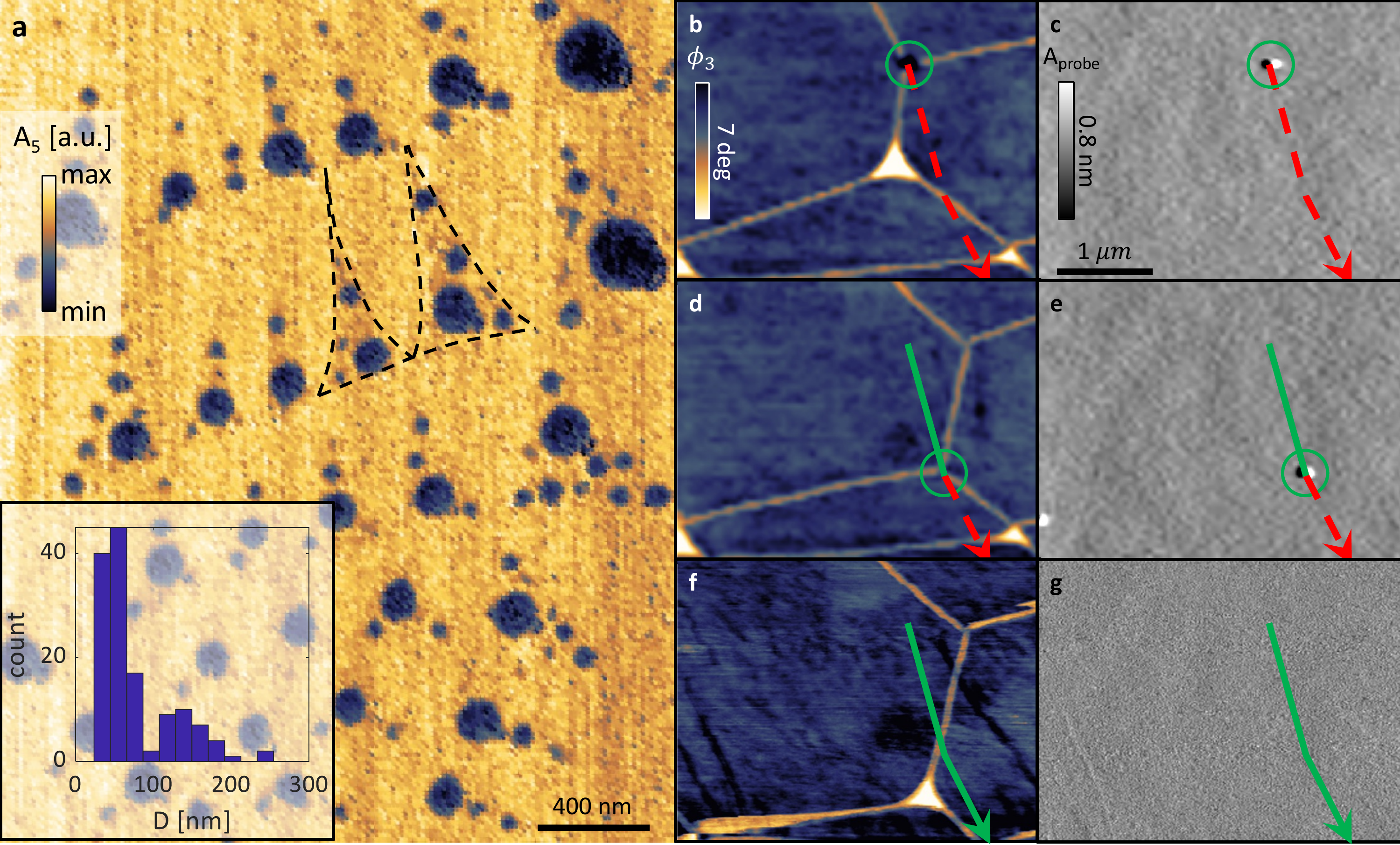}
    \caption{{\textbf{Moir\'e nanoengineering via metallic nanoparticles acting on the moir\'e superlattice} \textbf{(a)} Mid-IR nearfield imaging (amplitude, 5th harmonic) after ablation deposition of Fields metal on tDBG, resulting in the distribution of nano-particles. The metallic nano-particles appear as dark and densely cover the rhombohedral phase of the moir\'e superlattice (schematically outlined by dashed black contours). Inset: Diameter histogram of the nano-particles appearing in the scan. \textbf{(b)} Mid-IR nearfield phase and \textbf{(c)} AFM tapping-mode probe-amplitude maps taken simultaneously over a tDBG surface. A metallic nanoparticle was found at the intersection of three double domain walls (DDWs). The particle (circled in \textbf{(b,c)}) seemed to be confined by the trapping potential of the DDWs. \textbf{(d,e)} A similar measurement performed after pushing the particle down toward the rhombohedral domain (bright triangle in \textbf{(b)}) using an AFM tip in contact mode. The particle settled in a new position (circled in \textbf{(d,e)}) at the center of the rhombohedral domain. The rhombohedral domain appears to have collapsed to a point due to proximity of the particle. \textbf{(f)} Nearfield and \textbf{(g)} probe amplitude maps after pushing the particle away from the rhombohedral domain. The rhombohedral domain seems to have recovered following removal of the particle. The overlaid path indicates the trajectory of the particle, where the already traveled vs the future parts are indicated by solid-green and red-dashed lines respectively. All panels share the scale-bar in \textbf{(c)}. All nearfield panels share a color-scheme, and all topography panels share a color-scheme.} \label{Fig: backaction_by_nanoparticles}} 
    %\textbf{(e-h)} Elastic energy density maps from mechanical relaxation calculations of the formed domain at the presence of Gaussian perturbation (centered at white circle in \textbf{(e-h)}) to the stacking energy difference between the rhombohedral and Bernal domains. These calculations demonstrate how a local perturbation can lead to a global reconfiguration of the domain even up to a domain collapse \textbf{(g)}.
\end{figure*}

Having demonstrated moir\'e-assisted chemistry via domain-dependent surface interaction in twisted graphene heterostructures, we then explore whether this effect can be used to control the underlying moir\'e superlattice, for example, by mechanical manipulation of nanoparticles on the surface. Field's metal nanoparticles on a minimally-twisted tDBG sample demonstrate preferential adhesion to the rhombohedral phase, as shown in the mid-IR nearfield scan of Fig. \ref{Fig: backaction_by_nanoparticles}a. The nanoparticles form over and tile the rhombohedral domain in a characteristic pattern; a single larger nanoparticle forms in the center of the domain, surrounded by smaller nanoparticles that are all bounded by the rhombohedral domain. A histogram of the nanoparticle sizes after ablation deposition (Fig. \ref{Fig: backaction_by_nanoparticles}a inset) further demonstrates this bimodal distribution, with a majority of particles tending on the sub-100 nm scale.  The initial deposition of these nanoparticles on the graphene surface does not appear to affect the rhombohedral domain size. In contrast, in Fig. \ref{Fig: backaction_by_nanoparticles}b-g we demonstrate that mechanical manipulation of an individual nanoparticle can reversibly impact the local stacking configuration of the tDBG structure in its vicinity.

A nanoparticle at the intersection of three double domain walls (Fig. \ref{Fig: backaction_by_nanoparticles}b) is pushed by an AFM tip in contact mode towards a nearby rhombohedral domain (as indicated by arrow direction in Fig. \ref{Fig: backaction_by_nanoparticles}b). When the particle is displaced to the center of the rhombohedral domain, we observe that the domain has collapsed to a point around the location of the nanoparticle (Fig. \ref{Fig: backaction_by_nanoparticles}d-e). When the particle is pushed further away from the region, the rhombohedral domain appears to recover. Hence, by pushing a nanoparticle towards and away from the rhombohedral domain, we reversibly alter the size of the domain. We therefore demonstrate that the effect of domain-dependent adhesion is not passive but can actually cause a non-local reconfiguration of the lattice and the symmetry of the system by altering the stacking order around the nanoparticle. Our experimental results imply that the proximity of the nanoparticle appears to affect the stacking energy difference between rhombohedral and Bernal phases of tDBG. This causes a local perturbation, driving a local rearrangement of the domain to the extent of full collapse of the rhombohedral domain, which we expect to be reversible upon removal of the nanoparticle (as observed in Fig. \ref{Fig: backaction_by_nanoparticles} b-g). 

%The exact mechanism leading to this profound effect of metal nanoparticles over the moir\'e superlattice remains an open question that requires further investigation.

We demonstrate three robust examples of domain-dependent adhesion under varying conditions. While qualitative differences exist among these three demonstrations, we attribute these differences to the conditions under which the experiments were performed. The surface temperature of the graphene systems during metallic nanoparticle dispersion may affect the morphology of the dispersion, resulting in either a thin film (Fig. \ref{Fig: selective_adhesion_TDBG}) or discrete distributions of nanoparticles on the rhombohedral phase (Fig. \ref{Fig: backaction_by_nanoparticles}). In the case of water (Fig. \ref{Fig: selective_adhesion_TDTG}), we conjecture that the coverage of a certain domain will be additionally affected by the surface tension of the adsorbent. Thus, while we expect it would be possible to observe the domain-dependent adhesion of water and metal on both tDBG and tDTG systems, a clearer illustration of these phenomena would require the correct combination of surface temperature and adsorbent surface tension, as well as a more complete understanding of the mediating agent which is likely responsible for revealing domain-dependent adhesion.

% \section{Conclusions}

Our experiments reveal a domain-dependent surface chemistry with robust global effects. We observe preferential adherence of various materials to the rhombohedral domains in tDBG and rhombohedral-analogous domains in tDTG systems with finely resolved features that span large areas of the twisted heterostructures. Observation of samples both with and without domain-dependent adhesion, as well as the degradation of fine features in various environmental conditions, suggest that domain-dependent adhesion is not intrinsic to the materials studied. Our theoretical calculations indicate that domain-dependent adhesion is not an intrinsic phenomenon, which supports the suggestion that external mediating agents may play a role in the emergence of the domain dependence. Although the identity of this mediating agent remains an open question, we expect that it consists of an organic carbon-based substance to which the graphene systems are exposed during either the dry mechanical exfoliation of materials or the fabrication of the heterostructures (see Supporting Notes A and D).

 %The use of the moir\'e as a scaffold on which a periodic structure can be formed allows us the benefits and global degrees of freedom of the moir\'e to create a fine nanostructure of choice on a large scale without expensive fabrication techniques.

More importantly, our findings reveal that domain-dependent interaction is so significant that it strongly affects the configuration of both the adsorbates and the surface itself.
We observe that the local perturbation of a domain-preferentially adherent nanoparticle causes an extended, reversible reconfiguration of the moir\'e superlattice, namely, collapse and reappearance of a rhombohedral domain, around the manipulated nanoparticle. 
This allows us to propose a new approach, mechanical manipulation of the adsorbate, to reconfigure the underlying moir\'e superlattice. This additional level of control over stacking configuration may suggest a path towards additional control over the symmetry and electronic structure as well.

Further understanding of the domain-dependent adherent origin of nanoparticles could offer a process to manipulate systems of interest in-situ, a complementary capability to existing emerging techniques for the study and utilization of twisted heterostructure systems.~\cite{Gao2021heterostrain,qiao2018,huder2018,YankowitzChen2019,finney2019tunable} Control of such surface manipulation mechanisms could lead to a dynamic approach towards fine control of moir\'e structures and more generally, to moir\'e engineering. Additionally, investigation of domain-dependent adhesion in non-graphene moir\'e systems could be an avenue for future research. \cite {NatRevMat2021, Chiodini2022} In particular, moir\'e superlattices in transition metal dichalcogenide systems offer a rich platform for studying possible domain-dependent surface effects in stacking configurations not yet examined in our work. \cite{Wang2022, Weston2022}

%Overall, our work showcases how moiré systems can serve as a scaffold to generate fine nanostructure in large scale without expensive and/or scale-limited fabrication techniques, while allowing varying surface chemistry through selective adhesion, and remain controllable through nanoparticle manipulation. We believe that moiré scaffold, together with moiré chemistry and engineering elements holds great promise as a complete platform for future basic research and practical applications.

Overall, our work showcases how moir\'e systems can serve as a scaffold to generate fine nanostructures in large scale without expensive and/or scale-limited fabrication techniques. Additionally, we demonstrate a varying surface chemistry through selective adhesion that is also controllable through nanoparticle manipulation. We believe that the moir\'e scaffold, together with moir\'e chemistry and engineering elements, holds great promise as a complete platform for future basic research and practical applications.

\section{Supporting Information Available}
\begin{itemize}

    \item {Experimental methods}
    \item {First-principles calculations of the effect of various metals on ABAB and ABCA graphene}
    \item {Mid-IR nearfield phase and topography of pristine twisted double trilayer graphene}
    \item {Detailed history of the samples}
    \item {Molecular dynamics simulations and continuum
solvation model DFT calculations for effect of water on ABAB and ABCA graphene}
    \item {Demonstration of tDBG sample with and without moiré chemistry after surface treatment}
    \item {Demonstration of metallic-film coverage of the rhombohedral phase of tDBG at larger scales}
    
\end{itemize}

\section{Acknowledgments}
Nano-optical experiments at Columbia, including the efforts by V.H., D.H., C.R.D., and D.N.B., are supported as part of Programmable Quantum Materials, an Energy Frontier Research Center funded by the U.S. Department of Energy (DOE), Office of Science, Basic Energy Sciences (BES), under award DE-SC0019443. Research on moir\'e structures  at Columbia and Harvard is supported by ARO MURI: ARO (W911NF2120147). The development of nano-optical methods  is supported by the Vannevar Bush Faculty Fellowship ONR-VB: N00014-19-1-2630 (DNB). D.N.B. is Moore Investigator in Quantum Materials EPIQS GBMF9455. Work at Cornell was supported by the Cornell Center for Materials Research with funding from the NSF MRSEC program (DMR-1719875) and by the Gordon and Betty Moore Foundation’s EPiQS Initiative, Grant GBMF10436 to Eun-Ah Kim. D.H. is supported by a grant from the Simons Foundation (579913). M.P. is supported by the Swiss National Science Foundation (SNSF) through the Early Postdoc.Mobility program (Grant No.\ P2ELP2-191706). Experimental work at Columbia was carried out in part through the use of Columbia Nano Initiative Nanofabrication Clean Room facilities and with the help of CNI Clean Room staff.
The work of E.K., Z.Z., M.P., G.R.S., M.A. and E.K. is supported in part by the STC Center for Integrated Quantum Materials, NSF Grant No. DMR-1231319 and the NSF Award No. DMR-1922172. %, and the ARO MURI grant W911NF-21-2-0147.

\bibliography{References.bib} 

%\clearpage
%\newpage

%\section{For Table of Contents Only}
%\begin{figure} 
%    \centering
%    \includegraphics [width = 17cm] {TOCgraphic_v2.pdf}
%    \caption{TOC graphic}
%    \label{TOCgraphic}
%\end{figure}

% \include{supp}

\end{document}

\section{Supporting Note A. Experimental methods}

\emph{Fabrication of twisted double bilayer graphene and twisted double trilayer graphene samples.} We exfoliate graphene and hexagonal boron nitride (hBN) flakes onto a silicon substrate with 285 nm of silicon dioxide grown on top. Using an optical microscope, we then identify suitable hBN flakes as well as bilayer and trilayer graphene flakes, the thicknesses of which we verify using the Green function on the microscope. Samples are then assembled using the standard dry transfer method using a slide with polycarbonate (PC) film on top of a polydimethyl siloxane (PDMS) dome. \cite{Wang2013} 

We use the slide to first pick up a BN flake of thickness 30-40 nm, which is then placed into contact with one half of the desired bilayer or trilayer graphene flake. After picking up the first half, we then rotationally misalign the transfer stage by a small angle of less than 1 degree before picking up the remaining half of the graphene flake using the tear and stack method \cite{KimYankowitz2016,Pizzocchero2016}. The completed heterostructure is then removed from the slide and placed onto a silicon dioxide chip for subsequent characterizations and measurements, with the twisted graphene face exposed and the polymer film between the BN and silicon dioxide.

\smallskip
\emph{Mid-IR nearfield imaging and atomic force microscopy of rhombohedral and Bernal-stacked twisted double bilayer graphene domains.} The mid-IR nearfield scans in this work were acquired with a phase-resolved scattering type scanning optical microscope imaging (s-SNOM) with a commercial system (Neaspec), using a mid-IR quantum cascade laser (Hedgehog by Daylight Solutions) tuned between $8.7-10.2~\mu m$. The laser light was focused to a diffraction limited spot at the apex of a metallic tip, while raster scanning the sample at tapping mode. We collect the scattered light (power of $3-5~m$W) by a cryogenic HgCdTe detector (Kolmar Technologies). The nearfield amplitude and phase were extracted as harmonic components of the tapping frequency using an interferometric detection method, the pseudo-heterodyne scheme, by interfering the scattered light with a modulated reference arm at the detector\cite{Sunku2018h}. The sample topography was simultaneously acquired by tapping mode AFM. Dedicated AFM topography scans (without nearfield imaging) were performed on a Bruker Dimension Icon with a Nanoscope V Controller.

\smallskip
\emph{Mechanical ablation of Field’s metal on exposed twisted double bilayer graphene.} Placing the samples on top of a thermal stage, we melt a bulk piece of Field’s metal at an elevated temperature above the melting point of Field's metal (at 62$~^\circ C$). We use a tungsten tip to manually pull a narrow contact from the bulk Field’s metal. The contact is then placed within 50 um of the exposed graphene face. We pull the contact parallel to the longer orientation axis, aiming to break the metal at the point closest to the graphene sample in order to scatter nanoparticles on the exposed surface. By controlling the stage temperature one can affect the morphology of the formed film. The uniform film of Fig. 1 was formed at 76$~^\circ C$ while the nanoparticles formation of Fig. 3 was a result of an ablation at lower stage temperature a few degrees above the melting point.

\clearpage

\section{Supporting Note B. Effect of metal on tDBG: theory}
\subsection{Modeling effects of a metallic atom and water on the domain energy difference} 
To check the role of metal and water in selective adhesion, we performed first-principles Density Functional Theory (DFT) calculations to obtain the total energy and binding energy of rhombohedral and Bernal-stacked double bilayer graphene with gold atoms and water molecules on top. 
We shift the bottom two layers with respect to the top layer and place the gold atom directly on top of a carbon atom of the top layer. We optimize the separation between the gold atom and the DBG system by calculating the total energy as a function of the distance and finding the optimal separation. We then compare the total energy of the DBG+Au system with the DBG system without Au. If the Au atom is the key component, we expect the combined system to exhibit a larger energy difference between ABAB and ABCA such that in a relaxed system, the ABCA domain would shrink. Bernal stacking is the lowest energy configuration. Without Au, the energy difference between ABAB and ABCA is 38 meV/unit cell, and with Au, it is 25 meV/unit cell. That is, the energy difference shrinks with the gold atoms, which contradicts experimental observations. We checked the total energy of different Au densities by placing a single Au atom per $2\times2$ and $3\times3$ and obtained qualitatively similar results, suggesting that Au alone does not explain the observed domain collapse. 
We then proceed to check the role water and whether water selectively adheres to one domain versus the other. Similarly, we shift the bottom graphene layers with respect to the top two layers. We place a water molecule on top of a carbon atom of the top layer and optimize the water-graphene separation. 
The binding energy is defined as follows,
\begin{equation}
    E_\mathrm{binding} = E_\mathrm{H2O+Gr} - E_\mathrm{H2O} - E_\mathrm{Gr},
\end{equation}
where $E_\mathrm{H2O+Gr}$ is the total energy of the combined system, $E_\mathrm{H2O}$ is the energy of a single water molecule, and $E_\mathrm{Gr}$ is the energy of the double bilayer graphene layers (ABAB or ABCA). Table~\ref{table:dft_binding} shows the binding energy for 3 different water configurations on top of a $2\times2$ TBLD cell.

\begin{table}[ht!]
\begin{tabular}{cccc}
\cline{1-3}
\multicolumn{1}{|c|}{} & \multicolumn{1}{c|}{ABAB} & \multicolumn{1}{c|}{ABCA} &  \\ \cline{1-3}
\multicolumn{1}{|c|} {H2O configuration 1} & \multicolumn{1}{c|}{-0.481} & \multicolumn{1}{c|}{-0.493} &  \\ \cline{1-3}
\multicolumn{1}{|c|}{H2O configuration 2}  & \multicolumn{1}{c|}{-0.479} & \multicolumn{1}{c|}{-0.500} &  \\ \cline{1-3}
\multicolumn{1}{|c|}{H2O configuration 3}  & \multicolumn{1}{c|}{-0.474} & \multicolumn{1}{c|}{-0.495} &  \\ \cline{1-3}
                       &                       &                       & 
\end{tabular}
\caption{Binding energy (eV/unit cell) of a 2x2 DBG unit cell with different configurations of water on top}\label{table:dft_binding}
\end{table}

%The DFT calculations were performed using Vienna Ab Initio Simulation Package (VASP). We relax the electronic degrees of freedom to $1\times10^{-7}\,\mathrm{eV}$ and the ionic degrees of freedom to $1\times10^{-6}\,\mathrm{eV}$. We use $5\times5$ Monkhorts-Pack meshes centered at the $\Gamma $. We adopt the SCAN+rVV10 functionals to include van der Waals corrections. 

\begin{figure}[h!]
    \centering
    \includegraphics[width=\linewidth]{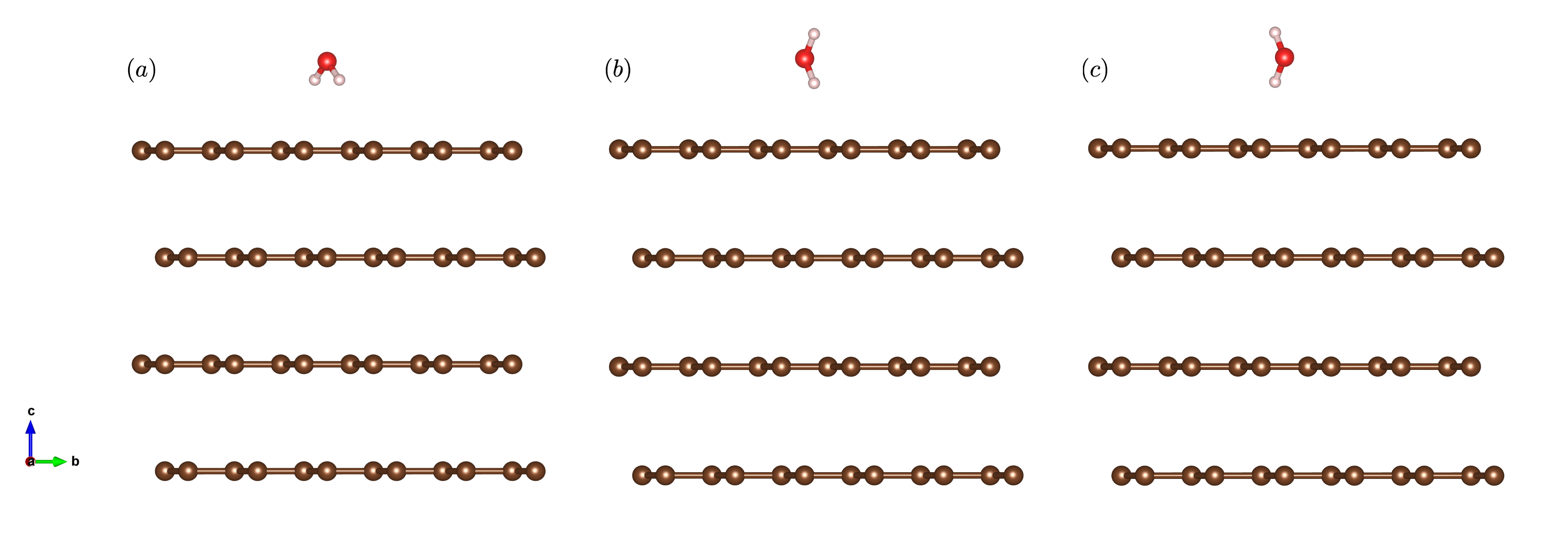}
    \caption{Side views of three types of water configuration. (a)-(c) correspond to the three configurations listed in Table~\ref{table:dft_binding} respectively.}
    \label{fig:config}
\end{figure}

\paragraph{Computational details}
%\subsection{Zoe} 
DFT calculations were performed using Vienna Ab Initio Simulation Package (VASP). We relax the electronic degrees of freedom to $1\times10^{-7}\,\mathrm{eV}$ and the ionic degrees of freedom to $1\times10^{-6}\,\mathrm{eV}$. We use $5\times5$ Monkhorts-Pack meshes centered at the $\Gamma $. We adopt the SCAN+rVV10 functionals to include van der Waals corrections. 

\subsection{Metal and water: effect of water on metallic work functions} 
\begin{figure}[h!]
    \centering
    \includegraphics[width=0.5\columnwidth]{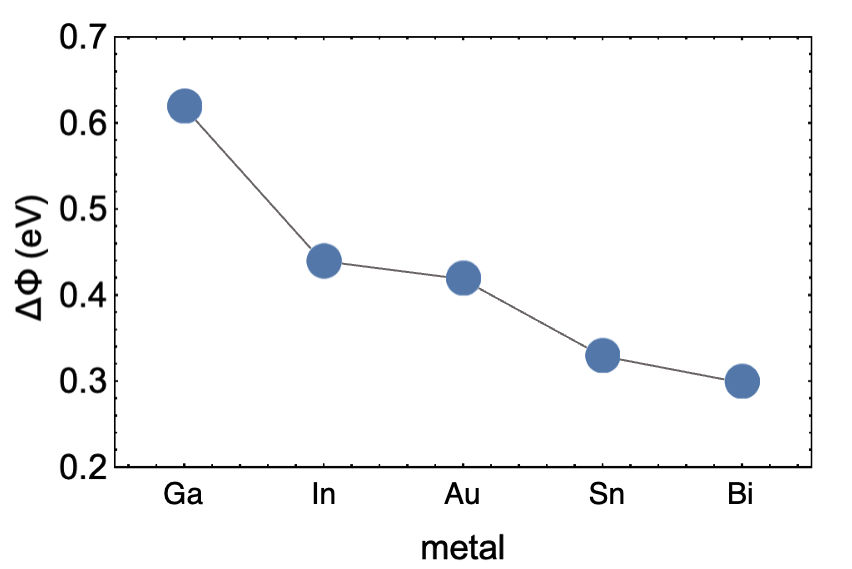}
    \caption{Difference $\Delta \Phi = \Phi_{vac}-\Phi_{H_{2}O}$ of {\em in vacuo} and hydrated work functions of gallium (Ga), indium (In), gold (Au), tin (Sn) and bismuth (Bi).}\label{FigWF}
\end{figure}

As a first step to investigating the interplay between water and the metal alloy, and their possible effects on the electronic properties of the graphene systems, we study the effect of water on the respective work functions of gallium (Ga), indium (In), gold (Au), tin (Sn) and bismuth (Bi). When the work functions of the two proximate systems (e.g., graphene and an alloy) differ, a potential step is formed at their interface, and charges may migrate across the barrier to equilibrate the Fermi levels when the two systems are brought into proximity. This can have substantial effects on the constituent systems' electronic properties; a Fermi level shift upwards (downwards) in system 1 with respect to neutrality means that electrons (holes) are donated by system 2 to system 1, making the latter $n$-type ($p$-type) doped. Since the magnitude of these shifts depends on the work function difference, if the presence of water impacts the systems' work functions asymmetrically, the electronic properties will also be affected. For each metal, we perform density-functional theory calculations on several metallic layers both {\em in vacuo} and in water, using the charge-asymmetric nonlocally determined local-electric (CANDLE) solvation model\cite{candle} to simulate the aqueous solution environment, as implemented in JDFTx\cite{Sundararaman2017SoftwareX}. The work functions in each environment are extracted from the difference in electrostatic potential inside and outside the metallic surface. The difference $\Delta \Phi = \Phi_{vac}-\Phi_{H_{2}O}$ of {\em in vacuo} and hydrated work functions of the elements comprising the metal alloy are shown in Fig. \ref{FigWF}. We find that for these elements, $\Delta \Phi$ ranges from 0.3 to 0.7 eV, which constitutes a significant modification of the potential at the surface. 

\paragraph{Computational details}
%\subsection{Eli} 
Our \textit{ab initio} density-functional theory (DFT) calculations were carried out within the total-energy plane wave density-functional pseudopotential approach, using Perdew-Burke-Ernzerhof generalized gradient approximation functionals\cite{Perdew1996PRL} and optimized norm-conserving Vanderbilt pseudopotentials in the SG15 family\cite{Schlipf2015Comp.Phys.Comms}, including van der Waals corrections\cite{vdWcorr}. Plane wave basis sets with energy cutoffs of 30 hartree were used to expand the electronic wave functions. We used fully periodic boundary conditions and a $6 \times 6 \times 1$ $k$-point mesh to sample the Brillouin zone. Electronic minimizations were carried out using the analytically continued functional approach starting with a LCAO initial guess within the DFT$++$ formalism\cite{Freysoldt2009PRB}, as implemented in the open-source code JDFTx\cite{Sundararaman2017SoftwareX} using direct minimization via the conjugate gradients algorithm\cite{Payne1992RMP}. All unit cells were constructed to be inversion symmetric about $z=0$ with a distance of $\sim 60$ bohr between periodic images of the atomic surface, using coulomb truncation to prevent image interaction.

\clearpage

\section {Supporting Note C. Mid-IR nearfield phase and topography of pristine twisted double trilayer graphene}

\begin{figure} [h!]
    \centering
    \includegraphics[width=7.62cm]{Figure 3_SI.pdf}
    \caption{\textbf{(a)} Moir\'e-assisted chemistry in twisted double trilayer graphene (a) For small twist angles the twisted double trilayer graphene system (tDTG) forms domains of ABABAB and BCBACA stacking configurations \cite{Halbertal_2022}. (b) Mid-IR nearfield phase (3rd demodulation harmonic) of the tDTG, revealing the moir\'e superlattice. (c) A featureless topography map of the pristine (before wetting) moir\'e superlattice covering the same scan window. The scan was taken simultaneously with (b).}
\end{figure}

\clearpage

\section{Supporting Note D. Detailed history of samples}
\emph{Figure 1: tDBG and domain-dependent metal adhesion}

We fabricate this tDBG sample and mechanically ablate a Field’s metal on the graphene surface as described in experimental methods (Supporting Note A). After 45 days of keeping the sample in vacuum, we perform tapping mode atomic force microscopy (AFM) and note that fine features started to degrade. We then transfer this sample to N$_2$ and repeat tapping mode AFM on the sample after 173 days and 302 days of storage in N$_2$. We note that fine features seem stable in N$_2$, as features that have not degraded after the initial 45 day storage time in vacuum are still robust after nearly a full year in N$_2$ storage.

\emph{Figure 2: tDTG and domain-dependent water adhesion}

We fabricate this tDTG sample as detailed in experimental methods (Supporting Note A) and image it with simultaneous mid-IR nearfield and topographic mapping. While moir\'e is visible in the mid-IR nearfield image, the topography of the surface is pristine, with no features corresponding to the moir\'e pattern. We then place this sample into a sealed beaker at room temperature, suspending it above deionized water for 90 minutes but not immersing the sample within water. In a subsequent simultaneous mid-IR nearfield and topographic measurement, we observe that the water has adhered to the sample surface in a pattern which corresponds to the rhombohedral domain.

\emph{Figure 3: Moir\'e nanoengineering via metallic nanoparticles on tDBG}

We fabricate this tDBG sample as detailed in experimental methods (Supporting Note A). We mechanically ablate Field’s metal at 65C across the sample surface and observe nanoparticles tessellating across the surface. Performing simultaneous mid-IR nearfield and topographic imaging, we note the rhombohedral domain-dependence of the nanoparticle adherence. The sample is then thermally annealed in vacuum leaving only sparse nano-particles on the surface. We are able to manipulate the size and shape of a rhombohedral domain by manipulating the position of a nanoparticle towards and away from this domain using a contact mode AFM tip. After demonstrating the ability to reconfigure the moir\'e superlattice, we vacuum anneal the sample to remove the Field’s metal and store the sample in vacuum.

\clearpage

\section{Supporting Note E. Effect of water on tDBG: theory}
To assess whether the selective wetting experimentally observed is intrinsic to twisted double bilayer graphene, we perform \emph{ab initio} molecular dynamics (MD) simulations and continuum solvation model density-functional theory (DFT) calculations. Given the impracticality of the first-principles treatment of \emph{twisted} double-bilayer graphene due to the large number of atoms in the unit cell, we resort to simplified models consisting of hydrated \emph{untwisted} ABAB- or ABCA-stacked double bilayer graphene. These models closely describe the large domains experimentally imaged and enable to single out the role of stacking on the superficial water adhesion.

\subsection{Molecular dynamics simulations} 
\begin{figure}[th!]
    \centering
    \includegraphics[width=0.3\columnwidth]{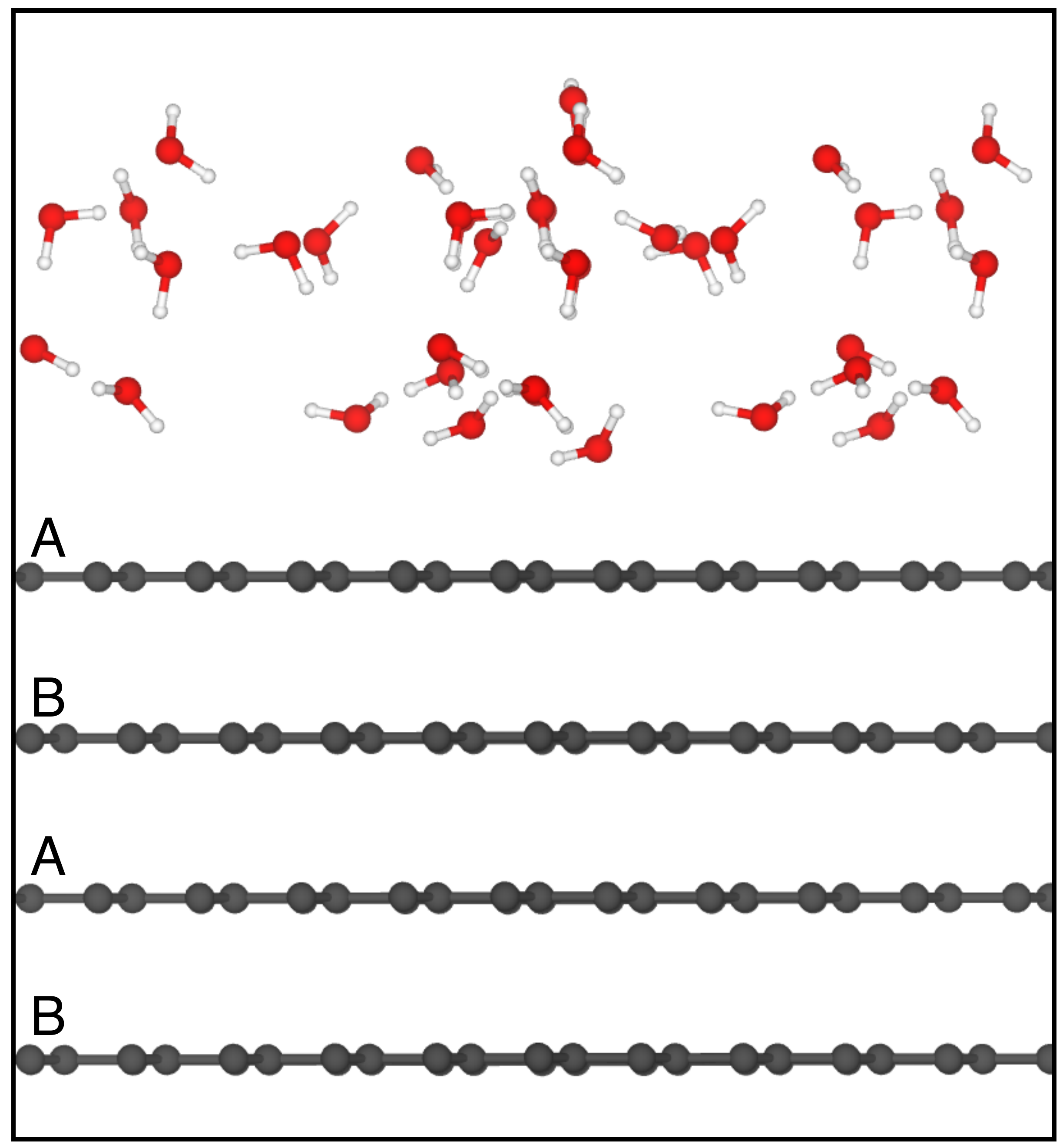}
    \caption{{Representative atomic structure of liquid water on four-layer graphene, consisting of ABAB-stacked four layer graphene in the (4 $\times$4)Gr$\cdot$16H$_2$O model.  Grey, red, and white balls represent carbon, oxygen, and hydrogen atoms, respectively. } \label{Fig1}}
\end{figure}

\begin{figure}[th!]
    \centering
    \includegraphics[width=0.8\columnwidth]{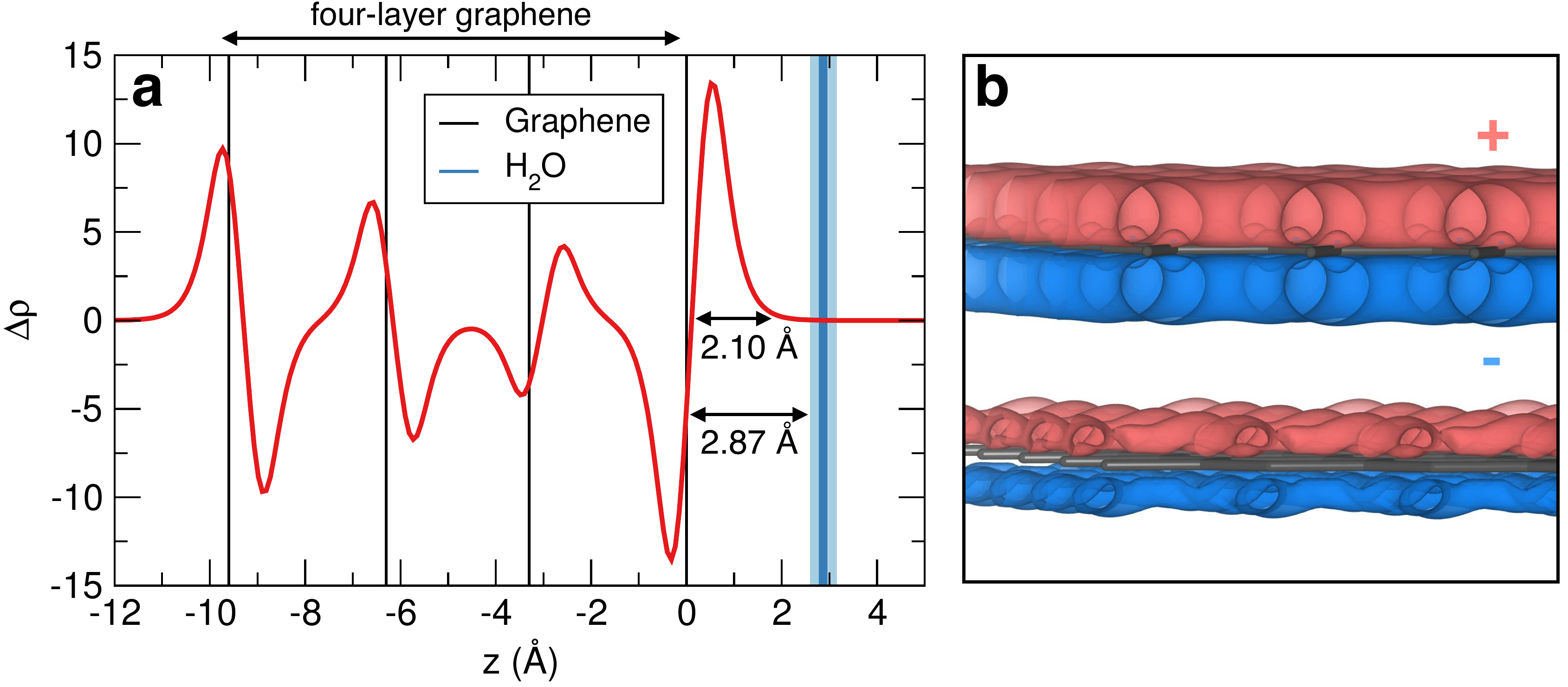}
    \caption{{(a) Difference in in-plane integrated charge density, $\Delta \rho$, between ABAB- and ABCA-stacked double bilayer graphene prior to wetting in the out-of-plane direction, $z$. The vertical black lines mark the position of each of the four layers; the position of the surface layer is set to zero. The vertical blue bar marks the average position of the water molecule that is the closest to the surface. (b) Difference in total charge density between ABAB- and ABCA-stacked double bilayer graphene prior to wetting. Only the two outermost layers are shown for clarity. Red and blue isosurface (0.005 $e$  {\AA}$^{-3}$) represent positive and negative contributions to the charge density, respectively. } \label{Fig2}}
\end{figure}

For each of the two stacking configurations, we consider two models of increasing sizes, i.e., (i) a (2 $\times$2) supercell of double bilayer graphene and four water molecules containing 44 atoms, which we refer to as (2 $\times$2)Gr${\cdot}$4H$_2$O, and (ii) a (4 $\times$4) supercell of double bilayer graphene and 16 water molecules containing 172 atoms, which we refer to as (4 $\times$4)Gr$\cdot$16H$_2$O. An illustrative example of our models is shown in Supporting Figure \ref{Fig1}. Supporting Table S1 lists the time-averaged adhesion energies for each of the four models investigated. These quantities are well-converged with respect to the size of our models, thus making us confident with the methodological approach adopted. Importantly, the adhesion energies ($\sim$23 meV {\AA}$^{-2}$) are found to be insensitive to the stacking configuration, hence indicating that the selective wetting experimentally observed is not intrinsically driven by the distinct stacking configurations that emerge in twisted double bilayer graphene. To understand the origin of the stacking-independent adhesion energies, we have inspected the charge density of double bilayer graphene prior to wetting. Specifically, we have integrated along the in-plane direction the charge density of ABAB- and ABCA-stacked configurations and visualized their difference, as shown in Supporting Figure \ref{Fig2}. The appreciable dissimilarity of the charge density between the two stacking configurations extends up to 2.10 {\AA} away from the surface layer. This value must be compared with the average distance between the surface layer and the closest water molecule. In agreement with an earlier work \cite{Brandenburg2019a}, these latter distances are found to be 2.87 {\AA} $\pm$ 0.17 and 2.87 {\AA} $\pm$ 0.19 for liquid water forming on top of  ABAB- or ABCA-stacked four-layer graphene, respectively. We thus conclude that liquid water is too distant from the surface to experience any distinction in electrostatics between the two stacking configurations of twisted double bilayer graphene.

\begin{table}[h]
\begin{tabular}{llll}
\cline{1-3}
\multicolumn{1}{|l|}{} & \multicolumn{1}{l|}{ABAB} & \multicolumn{1}{l|}{ABCA} &  \\ \cline{1-3}
\multicolumn{1}{|l|}{(2 $\times$2)Gr${\cdot}$4H$_2$O} & \multicolumn{1}{l|}{$-23.0 \pm 1.5$} & \multicolumn{1}{l|}{$-22.8 \pm 1.9$} &  \\ \cline{1-3}
\multicolumn{1}{|l|}{(4 $\times$4)Gr$\cdot$16H$_2$O} & \multicolumn{1}{l|}{$-23.1 \pm 2.1$} & \multicolumn{1}{l|}{$-23.6 \pm 1.5$} &  \\ \cline{1-3}
                       &                       &                       & 
\end{tabular}
\caption{Adhesion energy, in units of meV {\AA}$^{-2}$, of liquid water to ABAB- and ABCA-stacked four-layer graphene as obtained from \emph{ab initio} molecular dynamics simulations on the two models (2 $\times$2)Gr${\cdot}$4H$_2$O and (4 $\times$4)Gr$\cdot$16H$_2$O  discussed in the text.}
\end{table}

\paragraph{Computational details}
%\subsection{Michele}
%To assess whether the selective wetting experimentally observed is intrinsic to twisted double bilayer graphene, we perform \emph{ab initio} molecular dynamics simulations. 
Our calculations are carried out in the density-functional theory framework \cite{DFT}, as implemented in \textsc{vasp} \cite{VASP1, VASP2, VASP3, PAW}. We use the meta-generalized gradient approximation-based, van der Waals-inclusive SCAN-rVV10 exchange-correlation functional \cite{Peng2016}, which was shown to provide an accurate description of neat liquid water \cite{Wiktor2017}.
%Given the impracticality to achieve a first-principles treatment of \emph{twisted} double-bilayer graphene due to the large number of atoms in the unit cell, we resort to simplified models consisting of hydrated \emph{untwisted} ABAB- or ABCA-stacked double bilayer graphene. These models closely describe the large domains experimentally imaged and enable to single out the role of stacking, if any, on the superficial water adhesion. Specifically, for each of the two stacking configurations, we consider two models of increasing sizes, i.e., (i) a (2 $\times$2) supercell of double bilayer graphene and 4 water molecules containing 44 atoms, which we refer to as (2 $\times$2)Gr${\cdot}$4H$_2$O, and (ii) a (4 $\times$4) supercell of double bilayer graphene  and 16 water molecules containing 172 atoms, which we refer to as (4 $\times$4)Gr$\cdot$16H$_2$O. An illustrative example of our models is shown in Supporting Figure \ref{Fig1}. The cutoff on kinetic energy is set to 500 eV and the Brillouin zone is sampled with a mesh 6 $\times$ 6 and 2 $\times$ 2 $k$-points for the (2 $\times$2)Gr${\cdot}$4H$_2$O and (4 $\times$4)Gr$\cdot$16H$_2$O models, respectively.
Molecular dynamics simulations are performed according to the Born-Oppenheimer scheme within the canonical ($NVT$) ensemble where the temperature is maintained at $T = 350 $ K. We have replaced hydrogen atoms with deuterium atoms and integrated the equations of motion with the Verlet algorithm along with a timestep of 0.5 fs. Our analysis is conducted on molecular dynamics runs with a duration of 15 and 10 ps for the (2 $\times$2)Gr${\cdot}$4H$_2$O and (4 $\times$4)Gr$\cdot$16H$_2$O models, respectively, which are preceded by equilibrium runs of 10 and 5 ps. To evaluate the strength of the interaction between liquid water and ABAB- or ABCA-stacked double bilayer graphene, we have determined the adhesion energy, $E\textsubscript{A}$. This is accomplished by selecting, for each trajectory, a snapshot every 0.25 ps and obtaining the instantaneous $E\textsubscript{A}$ through single-point calculations as
\begin{equation}
E\textsubscript{A} = \frac{E\textsubscript{W+G} - (E\textsubscript{W} + E\textsubscript{G})}{A},
\end{equation}
where $E\textsubscript{W+G}$ is the total energy of the system, $E\textsubscript{W}$ and $E\textsubscript{G}$ are the total energies of double bilayer graphene and liquid water, respectively, $A$ is the area of the unit cell, $A = (na_0)^2\sin(\frac{\pi}{3})$, with $n$ being the size of the supercell ($n = 2$ or $4$) and $a_0$ the lattice constant of graphene. According to this expression, negative values of $E\textsubscript{A}$ denote exoergic processes. 

\subsection{Continuum solvation model} 
\begin{figure}[h!]
    \centering
    \includegraphics[height=0.4\columnwidth]{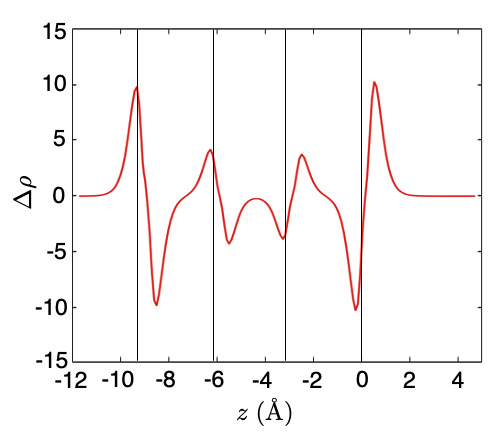}
    \caption{Difference in planar-integrated charge density $\Delta \rho$ between ABAB- and ABCA-stacked double bilayer graphene in the normal ($z$) direction, in an aqueous environment. The vertical black lines mark the position of the four graphene layers, the position of the surface layer is set to zero.} \label{FigElecDensJDFTx}
\end{figure}

We further study the effect of wetting on the distinct stacking configurations present in twisted double bilayer graphene (tDBG), using the charge-asymmetric nonlocally determined local-electric (CANDLE) solvation model\cite{candle} to simulate the electrostatics of graphene multilayers in an aqueous solution environment, as implemented in JDFTx\cite{Sundararaman2017SoftwareX}. To this end, we calculate the charge density of hydrated ABAB- and ABCA-stacked double bilayer graphene (DBG) separately and compute their difference, whose planar-integrated value is shown in Figure \ref{FigElecDensJDFTx}. That a marked difference in surface electrostatics between the two stacking configurations of DBG persists under the dielectric effects of water is evident from our continuum solvation calculations, which are consistent with the MD simulation results. These results suggest the electrostatic difference between stackings endures in the case of both dry and wet systems, providing further support for the hypothesis that the experimentally observed selective wetting is not intrinsic to tDBG.

\paragraph{Computational details}
%\subsection{Eli} 
Our \textit{ab initio} density-functional theory (DFT) calculations were carried out within the total-energy plane wave density-functional pseudopotential approach, using Perdew-Burke-Ernzerhof generalized gradient approximation functionals\cite{Perdew1996PRL} and optimized norm-conserving Vanderbilt pseudopotentials in the SG15 family\cite{Schlipf2015Comp.Phys.Comms}, including van der Waals corrections\cite{vdWcorr}. Plane wave basis sets with energy cutoffs of 30 hartree were used to expand the electronic wave functions. We used fully periodic boundary conditions and a $6 \times 6 \times 1$ $k$-point mesh to sample the Brillouin zone. Electronic minimizations were carried out using the analytically continued functional approach starting with a LCAO initial guess within the DFT$++$ formalism\cite{Freysoldt2009PRB}, as implemented in the open-source code JDFTx\cite{Sundararaman2017SoftwareX} using direct minimization via the conjugate gradients algorithm\cite{Payne1992RMP}. All unit cells were constructed to be inversion symmetric about $z=0$ with a distance of $\sim 60$ bohr between periodic images of the atomic surface, using coulomb truncation to prevent image interaction.

\clearpage

\section {Supporting Note F. Demonstration of moir\'e-assisted chemistry switching with surface treatment}

%%% Formerly FIG 2
\begin{figure}[h!]
    \centering
    \includegraphics[width=7.62cm]{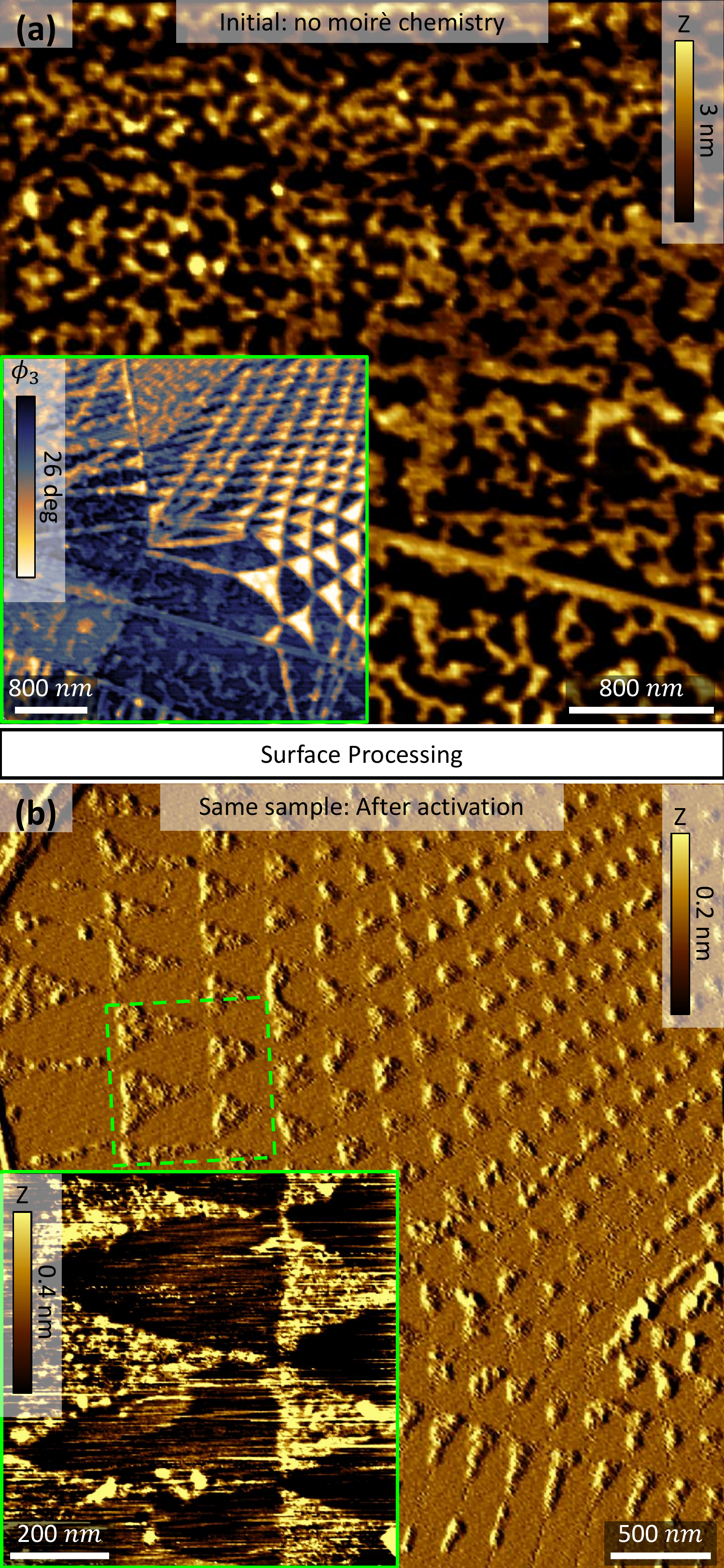}
    \caption{{\textbf{Moir\'e-assisted chemistry switching} \textbf{(a)} Topography of a tDBG surface after mechanical ablation deposition of Field's metal (see Methods) showing no correlation of metal patches with the moir\'e superlattice (inset - simultaneously taken mid-IR nearfield map). \textbf{(b)} Topography mapping (probe amplitude of tapping mode) of the same sample after moir\'e-assisted chemistry has been activated, showing strong adhesion of an unspecified residue to the rhombohedral phase (as highlighted in topography map in inset).} \label{Fig: selective_adhesion_switching}}
\end{figure}

We fabricate this tDBG sample as described in experimental methods (Supporting Note A). This sample undergoes vacuum thermal annealing to remove the polymer which was used to assemble the heterostructure, and additionally undergoes lithographic steps of electron-beam lithography, inductively coupled plasma etching, and electron beam gold deposition. After rinsing the sample with acetone to remove the PMMA polymer resist used in lithographic steps, we perform mid-IR nearfield imaging to confirm the presence of moir\'e.

We perform Field’s metal ablation deposition at 76\degree C, then vacuum thermal anneal the sample to remove the Field’s metal before repeating mid-IR nearfield imaging to confirm the moir\'e is unaffected. The tapping mode AFM scan of Fig. \ref{Fig: selective_adhesion_switching}a shows patches of film covering the exposed tDBG surface without any correlation with the moir\'e superlattice (revealed by the simultaneously taken nearfield map shown in the inset).

We then dip the sample in deionized water for 2 minutes and dry the sample with N$_2$ flow. We repeat the mechanical ablation deposition of Field’s metal at 74\degree C and vacuum thermal anneal to remove the metal. We expose the sample to a humid environment while we repeat the AFM measurement in tapping mode. We observe the emergence of selective coverage of the rhombohedral domain (Fig. \ref{Fig: selective_adhesion_switching}b) corresponding with the rhombohedral domains seen in mid-IR nearfield imaging (inset). As with the sample shown in Fig. 1, the domain-dependent coverage shows fine features such as sharp corners with a $\sim$10 nm fidelity. A qualitatively similar pattern can be seen in an application note of an Kelvin probe force microscopy mode released by Asylum Research\cite{Asylum2021}. There, in Fig. 3b-c some unknown contamination selectively covers the ABC rhombohedral domain in trilayer graphene. 

Further manipulating the surface of this sample using the AFM in contact mode results in pushing of the selectively adhering substance to the side, exhibiting behavior unlike what is expected of water. We note that the behavior of this surface contaminant is similar to the rhombohedral domain-dependent behavior demonstrated in Fig. 3c-d of the Asylum Research application note \cite{Asylum2021}, which is noted as an unknown organic substance.

\clearpage

\section{Supporting Note G. Metallic film coverage of the rhombohedral phase of tDBG on a large scale}
\begin{figure*}[h!]
    \centering
    \includegraphics[width=15.24cm]{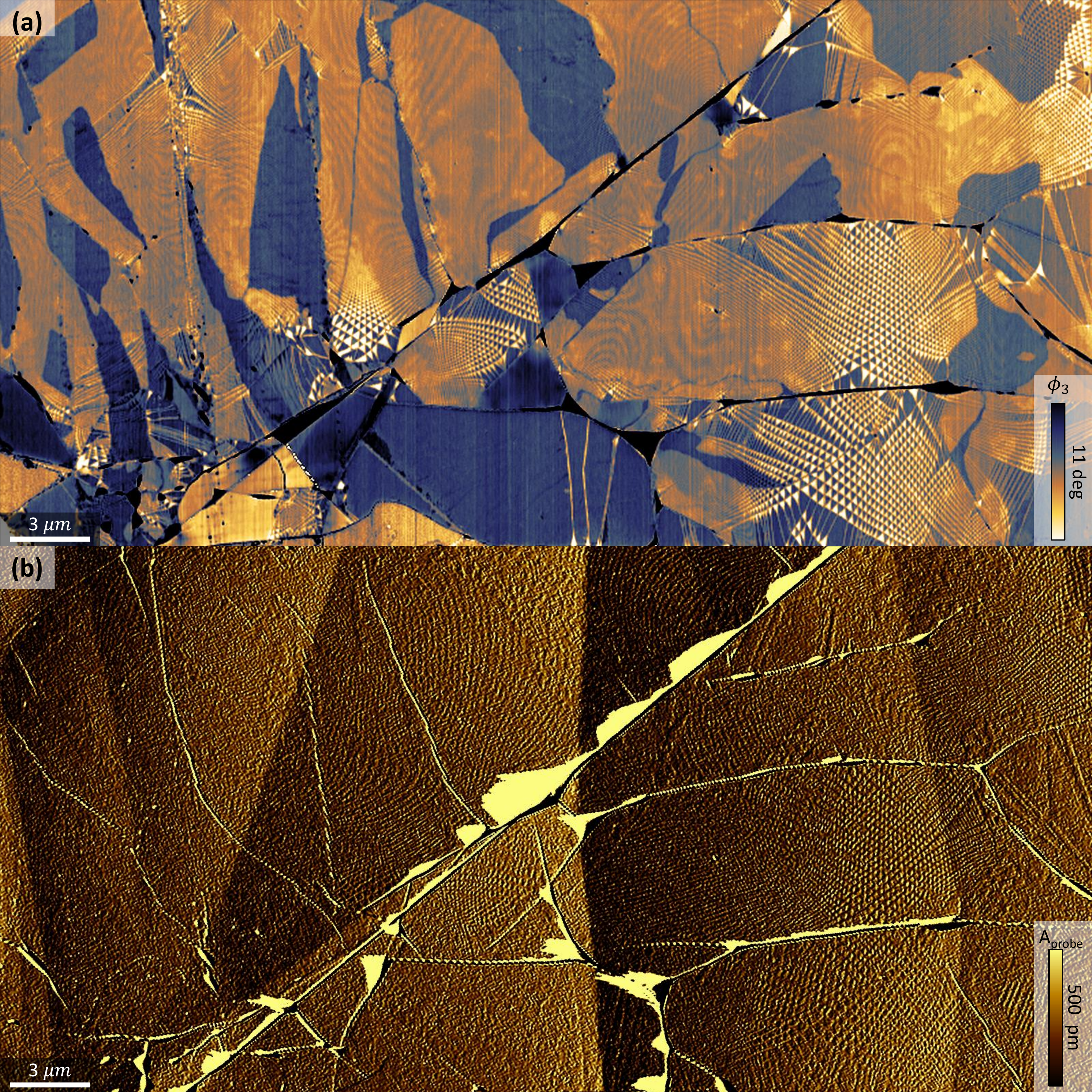}
    \caption{{\textbf{Moir\'e-assisted chemistry in twisted double bilayer graphene - a large scale scan} \textbf{(a)} Mid-IR nearfield imaging of the twisted double bilayer moir\'e superlattice (tDBG) moir\'e superlattice (phase, 3th harmonic). Image taken on pristine sample (same sample as Fig. 1) before deposition. No topographic feature were observed. \textbf{(b)} Metal nano-structures forming over the rhombohedral phase of tDBG, revealed by tapping mode atomic force microscopy (AFM). The plotted channel is the probe-amplitude (as it is less sensitive to large-scale z modulation).} \label{Fig: selective_adhesion_TDBG_large_scale}}
\end{figure*}

\clearpage

\bibliography{References.bib}